# Coupled stack-up volume RF coils for low-field MR imaging

Yunkun Zhao[1], Aditya A Bhosale[1], Xiaoliang Zhang[1,2*]

[1]Department of Biomedical Engineering, [2]Department of Electrical Engineering, State University of New York at Buffalo, Buffalo, NY, United States

*Corresponding author:

Xiaoliang Zhang, Ph.D.
Department of Biomedical Engineering
State University of New York at Buffalo
Bonner Hall 215E
Buffalo, NY, 14226
U.S.A.

Email: xzhang89@buffalo.edu

**Abstract**

The advent of low field open magnetic resonance imaging (MRI) systems has greatly expanded the accessibility of MRI technology to meet a wide range of patient needs. However, the inherent challenges of low-field MRI, such as limited signal-to-noise ratios and limited availability of dedicated RF coil, have prompted the need for innovative coil designs that can improve imaging quality and diagnostic capabilities. In response to these challenges, we introduce the coupled stack-up volume coil, a novel RF coil design that addresses the shortcomings of conventional birdcage in the context of low field open MRI. The proposed coupled stack-up volume coil design utilizes a unique architecture that optimizes both transmit/receive efficiency and RF field homogeneity and offers the advantage of a simple design and construction, making it a practical and feasible solution for low field MRI applications. This paper presents a comprehensive exploration of the theoretical framework, design considerations, and experimental validation of this innovative coil design. Through rigorous analysis and empirical testing, we demonstrate the superior performance of the coupled stack-up volume coil in achieving improved transmit/receive efficiency and more uniform magnetic field distribution compared to traditional birdcage coils.

**Introduction**

Magnetic resonance imaging (MRI) has evolved into an indispensable diagnostic tool, offering non-invasive and high-resolution visualization of internal anatomical structures and physiological processes within the human body [1-11]. While high-field MRI has demonstrated the significant signal-to-noise ratio (SNR) gain and dominated the field [12-22], low-field MRI (below 1 Tesla) [23-26], particularly those operating at the 0.5 Tesla (T) level, has garnered significant attention in recent years due to its unique advantages and clinical utility, as well as recent advances in artificial intelligence [27-30]. The appeal of low-field open MRI lies in its capacity to cater to a diverse patient population, including those with claustrophobia, obesity, and pediatric patients, who may find conventional closed-bore MRI systems challenging or uncomfortable.

The merits of low-field open MRI systems are clear. The spacious, open design of these systems not only addresses patient comfort and accessibility concerns but also facilitates a broader range of imaging scenarios, such as interventional procedures and imaging of larger anatomical regions. However, the transition to lower magnetic field strengths comes with its own set of challenges. One of the primary limitations is the inherently lower signal-to-noise ratio (SNR) [20, 31-34], which can compromise image resolution and hinder the detection of subtle anatomical or pathological details. Therefore, there exists a pressing need for innovative solutions that can harness the benefits of low-field open MRI systems while mitigating these inherent drawbacks.

Central to the success of any MRI system is the radiofrequency (RF) coil [35-47], a crucial component responsible for transmitting and receiving MR signals during the imaging process. The design and performance of the RF coil play a pivotal role in image quality, signal strength, and overall diagnostic accuracy [48-55]. In the context of low field open MRI, the current RF coil configurations face the challenge of limited RF field (B1 field) transmit/receive efficiency and field homogeneity, particularly along coil axis. To bridge this gap and harness the full potential of low field open MRI systems, in this work, we introduce a coupled stack-up volume coil, a novel RF coil design specifically tailored to address these challenges. To investigate and demonstrate the proposed design, we have taken the 0.5T as an example field strength and designed and constructed a prototype coupled stack-up volume coil operating in the 20MHz range. This innovative RF coil design can significantly improve RF field efficiency and also enhance the field homogeneity along the coil axis, ultimately elevating the performance of low-field open MRI systems. The proposed design of the coupled stack-up coil was analyzed using full-wave electromagnetic (EM) simulation and tested on the workbench with standard RF measurement procedures. The performance is further validated through a comparison study with a standard birdcage coil [50, 56].

**Methods**

*EM Simulation*

Figure 1 shows the layout of the coupled stack-up volume coil. The coupled stack-up volume coil design consists of a stack of seven identical and individual coils, meticulously arranged to create a cylindrical imaging area with dimensions of 300 mm in diameter and 300 mm in length. Each coil unit is equipped with a 60pF capacitance capacitor, carefully selected to optimize its resonance characteristics at the desired Larmor frequency. The coil is driven via the central coil element in this stack configuration, which provides efficient radio frequency (RF) signal transmission and reception throughout the imaging volume. The spacing between these individual coils has been carefully orchestrated, a critical design consideration aimed at achieving a high degree of field homogeneity within the imaging area. The circuit diagram and coil spacing is shown in the figure 1B. Based on the number of the coils, there are four resonant modes for the coupled stack-up coil and the lowest resonant mode is used for imaging because it has the highest field strength efficiency. A traditional birdcage coil with the same size of a coupled stack-up coil has also been built for comparison. In comparison study, a cylindrical air phantom of 200 mm in diameter and 300 mm in length and a dielectric constant of 1 was been placed in the center of the coils as an imaging area for field strength and distribution evaluation. Scattering parameters and B1 field efficiency map were used to evaluate the performance of the stacked coils in coupling and imaging. All magnetic and electric field plots were normalized to 1 W accepted power. Numerical results of the proposed designs were obtained using the electromagnetic simulation software CST Studio Suite (Dassault Systèmes, Paris, France).

*Bench Test Model Assembly*

Figure 2 shows photographs and dimensions of bench test models of the coupled stack-up volume coil and birdcage coil. The bench test models have the same dimensions as the simulation model. The electrical track of coupled stack-up volume coil was built using 6.35 mm width copper tape and mounted on a 3D-printed polylactide acid frame. The imaging resonant frequency was tuned to 21 MHz and matched to 50 ohms by careful selection of the capacitance value on each individual coil. We used 7 identical fixed tuning capacitors with 39 pF capacitance. The matching circuit was implemented as shown in Figure 2A. One capacitor with 330 pF connected in parallel to the feeding line was employed for impedance matching. Except for three coils located at the center and two sides, the remaining four coils are movable, and their position may be adjusted to achieve a homogenous B-field under different imaging objects. Most areas of the coil are hollow and can also be used to alleviate claustrophobia in patients.

For comparison, a low-pass birdcage coil has also been made. The birdcage coil model has the same dimensions as its simulation model and the coupled stack-up volume coil. It was built using 6.35 mm width thick copper tape on a cardboard structure. The birdcage coil has 8 legs with 8 tuning capacitors and was tuned to 21 MHz and matched to 50 ohms by tuning capacitors and a matching circuit.

*3-D Magnetic and Electric Field Mapping*

A sniffer positioning system combined with a magnetic and electric field measurement setup, shown in Figure 3, was used to visualize the B and E field distribution in the bench test. The system consists of a Genmitsu PROVerXL 4030 router (SainSmart, Lenexa, United States) as a positioning system, a Keysight E5061 Vector Network Analyzer (Keysight, Santa Rosa, United States) for data reception and analysis, and a B/E field sniffer to receive field strength data. The positioning system was programmed to measure the B or E field strength at a level above the coils with a data step of 0.5 mm. The design of the B and E field sniffers is also shown in Figure 3. The B field sniffer is a coaxial cable loop that can measure the magnetic flux flow through the center of the loop, and the E field sniffer is a coaxial cable with the outer conductor and medium removed at the tip. During the measurement, the coil assembly is connected to port 1 of the VNA and the sniffer is connected to port 2. The S21 value is recorded by the VNA and the final field strength is calculated using the following equation:

$$log(B) = \frac{1}{20} * (Pout - X - 20 * log(F)) \qquad (1)$$

Where *B* is the magnetic flux density in Tesla, *F* is the frequency of the received signal in megahertz, *Pout* is the probe output power into 50 ohms in dBm, and *X* is a scale factor from calibration. The calibration was taken place using the result from the magnetostatics method and finite-difference time-domain (FDTD) method on a 5cm diameter circular RF coil with one tuning and one matching capacitor and built with 16 AWG copper wire. Three calculation results were used, including the numerical solution and analytical solution of the Biot-Savart law:

$$B(r) = a_z \frac{\mu_0 I b^2}{2(z^2 + b^2)^{3/2}} \qquad (2)$$

The Biot-Savart law is used to find the magnetic flux density at a point on the axis of a circular loop of radius *b* that carries a direct current *I* to verify the magnetic field. The result from FDTD methods generated by the electromagnetic simulation model from simulation software CST Studio Suite has been used to verify electric measurement results. All three calculated and simulated results verified our measurement system is correct and accurate.

**Results**

*Simulated Resonant Frequency and Field Distribution*

Simulated scattering parameters versus frequency of the stacked coils are shown in Figure 4. As shown in the figure, strong coupling is created between the coils, resulting in split resonant peaks. Four resonant frequencies were generated, with the lowest frequency at 21 MHz and the highest at 37.6 MHz. In the matched circuit case, the fourth mode is not obvious due to the impedance matching to 21 MHz for the matched case. Figure 5 shows simulated Y-Z, X-Z, and X-Y plane B field efficiency maps inside phantom generated by coupled stack-up volume coils, in which both planes are at the center of the axis. A set of the multiple X-Y plane slices with different distances from the phantom center B field efficiency maps inside the phantom generated by coupled stack-up volume coil has also been shown. The simulation result shows the coupled stack-up volume coil has a great homogenous field which can be used for MR imaging.

*Measured Scattering Parameters and Field Distribution*

Figure 6A shows that the S-parameter vs. frequency plots of the coupled stack-up coil is in good agreement with the simulation results. Four resonant modes with 20.1 MHz, 28.2 MHz, 31.8 MHz, and 34.4 MHz were formed. Figure 6B shows the B field efficiency distribution map on Y-Z plane measured with 3-D magnetic field mapping system. Coupled stack-up volume coil shows significant homogeneity and strong B field efficiency on the Y-Z plane and is in accordance with the simulation result, which also indicates that the simulation results are accurate and reliable.

*Field Distribution and Efficiency Evaluation*

Figure 7 compares the simulated B1 field efficiency between the coupled stack-up coil and birdcage coil on three different planes with the B1 field efficiency distribution map. Table 1 also compares the standard deviation and average B1 field efficiency of the B1 field strength inside the phantom between the field generated by the coupled stack-up coil and birdcage coil. The result shows that the coupled stack-up coil has significantly higher B field efficiency and B field homogeneity compared with the birdcage coil. With an average of 9.6058 $\mu T/\sqrt{W}$ inside the phantom, the B field efficiency of the coupled stack-up volume coil is 47.7% higher than the average B field efficiency of birdcage coil. As for homogeneity, the standard deviation of B field generated by coupled stack-up volume coil is also 68% lower than birdcage coil.

Figure 8 compared the between B field efficiency of bench test model of coupled stack-up volume coil and birdcage coil. The measured B-field efficiency distribution is shown in Figures 8A and 8B. The measured magnetic field efficiency plot is consistent with the simulation results. Figure

8C and 8D shows the B field efficiency plot at the center line along the X-Z plane, Y-Z plane, and X-Y plane. Not only does the coupled stack-up volume coil have higher B field efficiency, but the rate of decreasing of the B field from the center to the sides of the birdcage coil is much higher. The B field efficiency of the coupled stack-up volume coil, with the highest field efficiency of 11.4835 µT/$\sqrt{W}$, only reduces by 11.4% when reaching the edge of the coil with a minimum value of 10.2 µT/$\sqrt{W}$. On the other side, the B field efficiency of the birdcage coil decreases by 49.2% from a maximum field efficiency of 7.3 µT/$\sqrt{W}$ at the center to a minimum of 3.7313 µT/$\sqrt{W}$ at two edges. The measured result validates that the coupled stacked coil has a strong and homogeneous field within the imaging area compared with the birdcage coil.

*Comparison with Equal Gap Stack-Up Coils*

To underscore the critical importance of the meticulously designed gap distances within the coupled stack-up volume coil design, we conducted a comparative simulation analysis with a hypothetical scenario employing equal gap stack-up coils. Figure 9A shows a coupled stack-up coil configuration in which all coils are evenly distributed along the length of the imaging area. The length of the coupled stack-up coil is still 300 mm but the distance between each coil is now 50 mm. Figure 9B shows the magnetic field efficiency distribution map in all three planes for the equal distance coil setup. As shown in the figure, the evenly distributed coupled stack-up coil demonstrates a non-uniform field distribution along coil axis, where the center of the coil has the highest magnetic field efficiency and decreases significantly as the distance from the center increases. Figure 8C compares the simulated B field efficiency along the coil center between the coupled stack-up volume coil and the equal distance gap coupled stack-up volume coil. The B field efficiency of the equal distance coupled stack-up volume coil decreases by 34.4% from its highest value of 12.17 µT/$\sqrt{W}$ to minimum value of 7.8 µT/$\sqrt{W}$ while the B field efficiency of coupled stack-up volume coil with careful distance designed only decreases by 17%. For comparison, the use of non-calculated, equal gap stack-up coils result in a less uniform magnetic field.

**Discussion and Conclusion**

Critical to the success of the coupled stack-up volume coil design is the meticulous arrangement of its individual coils or resonant elements. The magnetic field strength at each coronal plane within the phantom should be most affected by the coil closest to it. By moving the coupled coil closer to the edge of the phantom, where the B field strength is weaker, the local B field can be improved to match the B field strength at the center of the coil, thus improving the overall field homogeneity.

In the realm of low-field open MRI systems at 0.5 Tesla, the pursuit of enhanced image quality, diagnostic accuracy, and patient comfort has led to innovative approaches and technologies. This study has introduced the coupled stack-up volume coil, a novel radio frequency (RF) coil design engineered to address the challenges inherent in 0.5T open MRI systems, particularly with respect to transmit/reception efficiency and field homogeneity. Through a research framework encompassing electromagnetic simulations and benchtop characterizations, we have illuminated the substantial advantages offered by this innovative coil design.

Based on this coupled stack-up design, the size and setup can be changed to achieve imaging for other body parts, such as the chest, knees, limbs, or wrists. The gap between each individual coil inside the coupled stack-up coil can be adjusted to further improve the homogeneity of the B field inside different imaging objects depending on their properties.

In conclusion, the coupled stack-up volume coil is successfully designed, constructed and tested for low field MR imaging. The proposed work represents a transformative development in the field of low-field MRI, particularly open MRI. Its innovative design, carefully orchestrated coil arrangement, and optimized capacitance parameters converge to deliver a solution that outperforms the conventional birdcage coil in the aspects of B1 field efficiency and homogeneity. not only addresses the challenges posed by low-field MRI but also enhances its capabilities. The ability to achieve superior transmit/receive efficiency and field homogeneity positions this design as a promising avenue for advancing low-field MRI's diagnostic precision and clinical utility.

**Acknowledgments**

This work is supported in part by the NIH under a BRP grant U01 EB023829 and by the State University of New York (SUNY) under SUNY Empire Innovation Professorship Award.

**Figures**

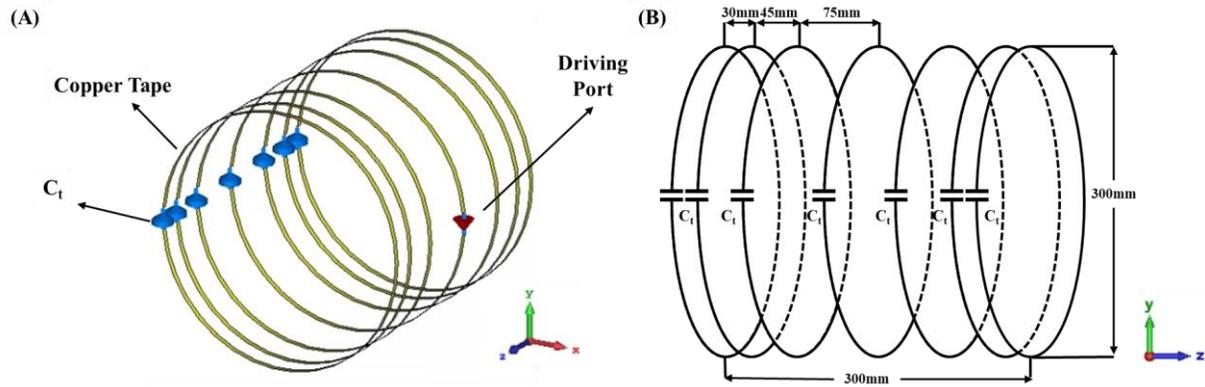

Figure 1
(A) Simulation model of coupled stack-up volume coil. (B) Circuit diagram of coupled stack-up volume coil. The distance between individual coils has been labeled in the figure.

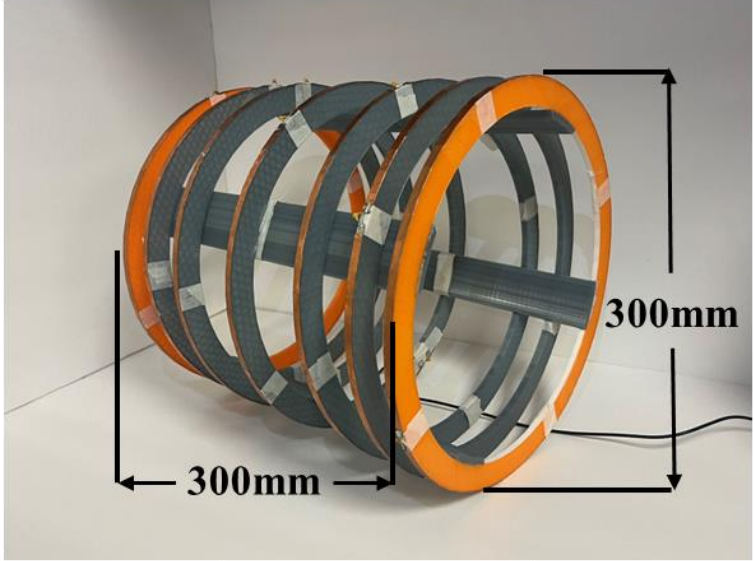
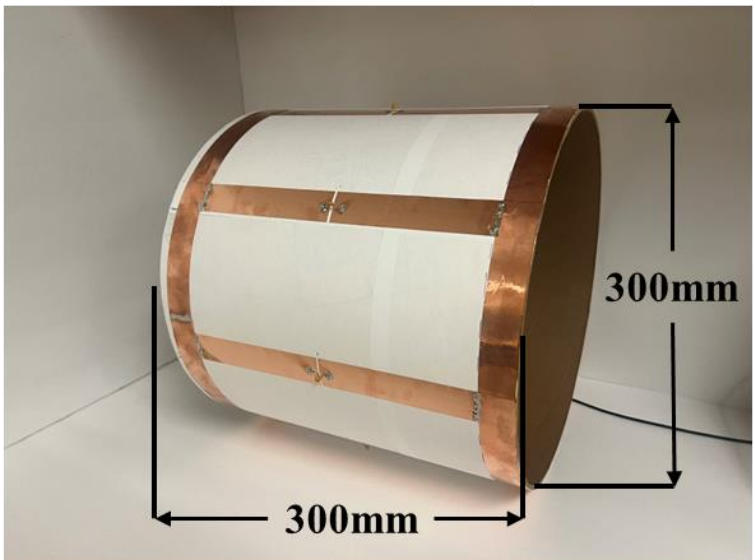

Figure 2
A photograph (A) of the bench test coupled stack-up volume coil model for imaging at 0.5T, corresponding resonant frequency of 21 MHz. For comparison, a custom-built 21 MHz low-pass birdcage coil (B) was used in this paper.

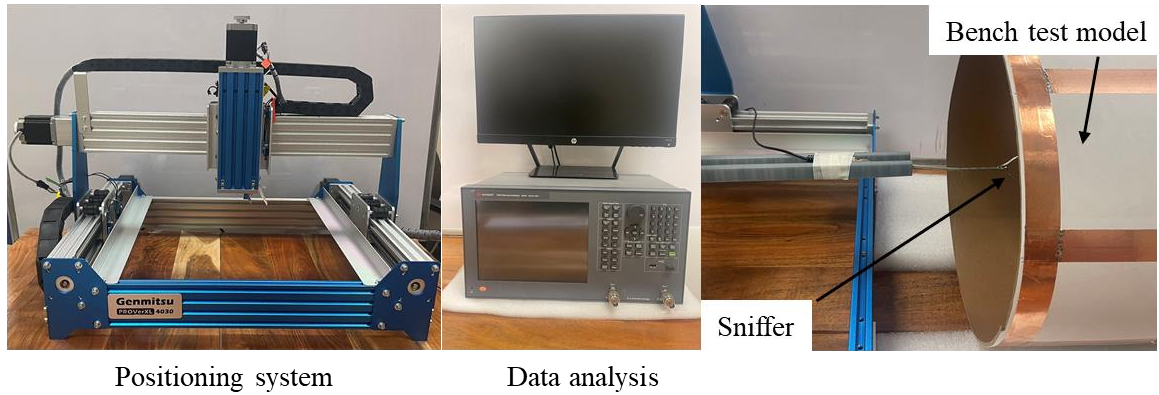

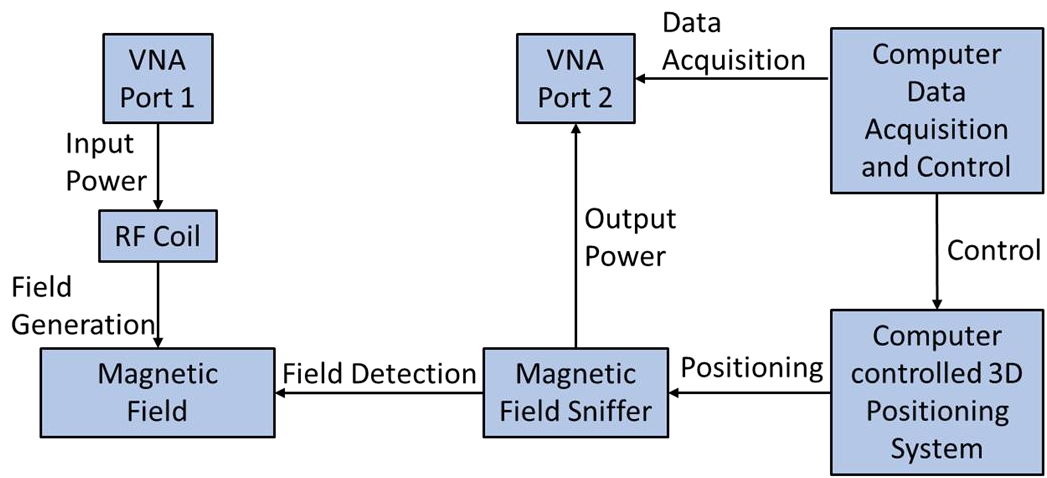

Figure 3
(A) Experimental setup of the sniffer-positioning system combined magnetic field measurement for the planar coupled array. The field of view (FOV) of the measuring system is 200 mm * 150 mm * 80 mm and the resolution is 0.5 mm * 0.5 mm. (B) Data processing flow for the 3-D magnetic field mapping system.

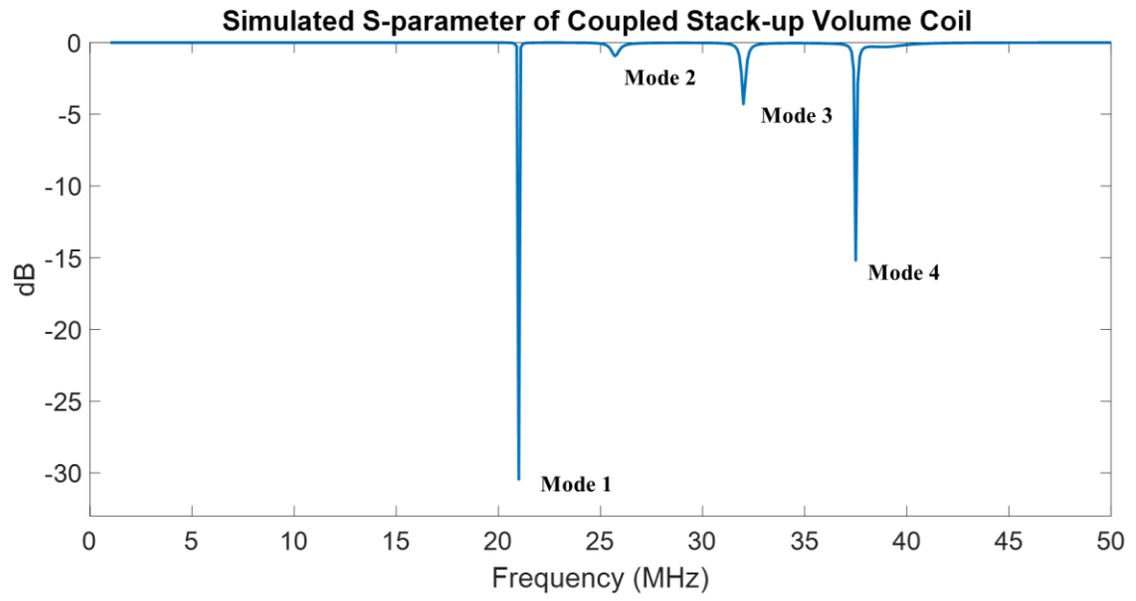

Figure 4
Simulated scattering parameters vs. frequency of the coupled stack-up volume coils.

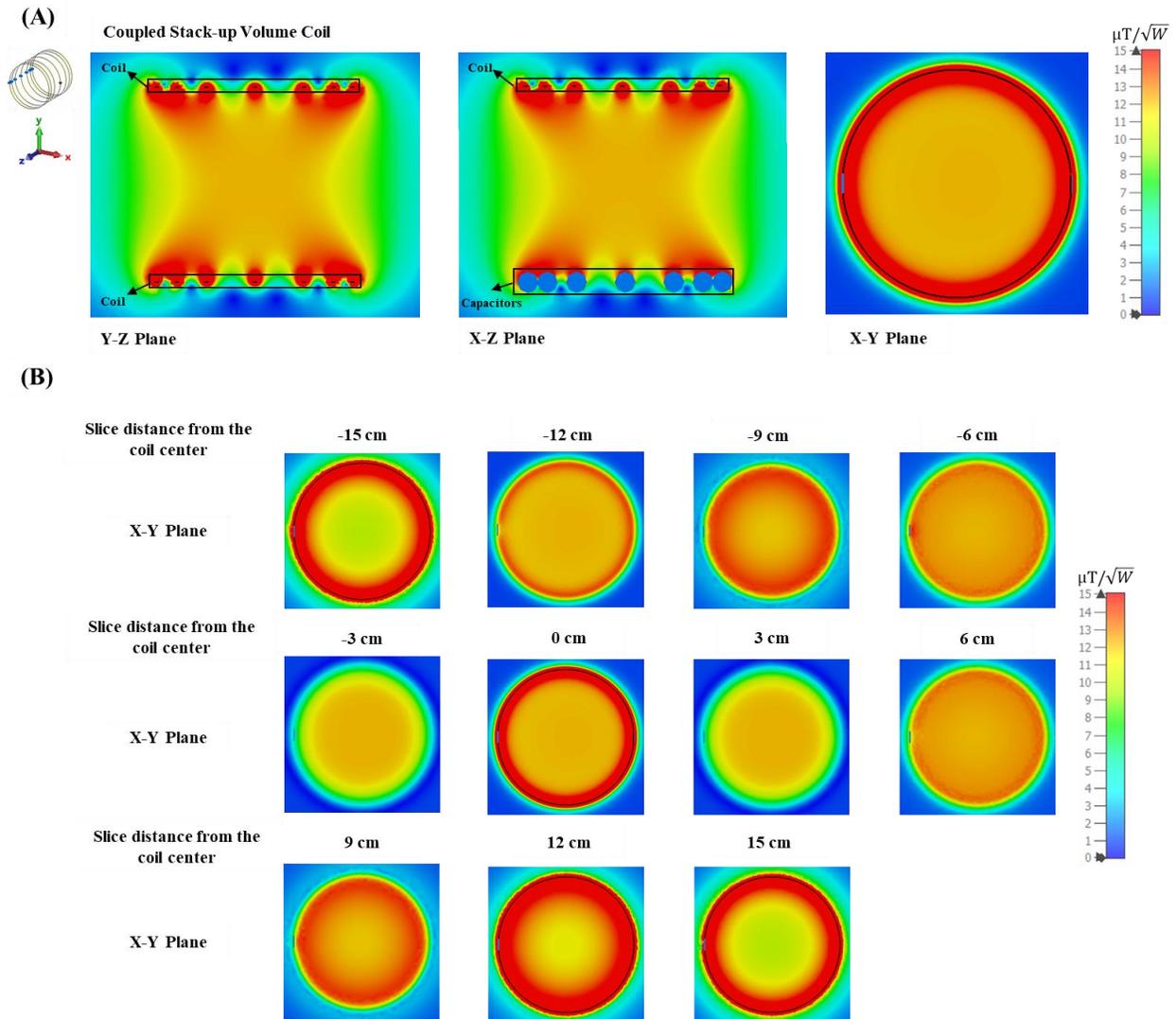

Figure 5
(A) Simulated Y-Z, X-Z, and X-Y plane B field efficiency maps inside phantom generated by coupled stack-up volume coils. Both planes are at the center of the axis. (B) A set of the multiple X-Y plane slices with different distances from the phantom center B field efficiency maps inside the phantom generated by coupled stack-up volume coil.

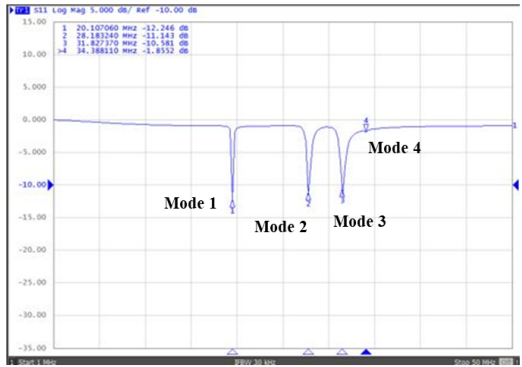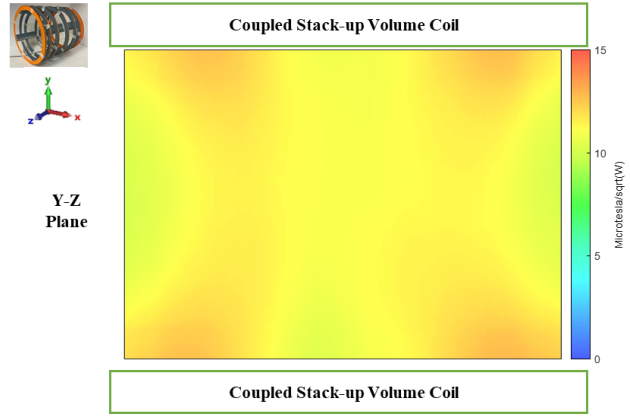

Figure 6
(A) Scattering parameters vs. frequency of the bench test model of coupled stack-up volume coils. (B) Measured B field efficiency maps on the Y-Z plane of coupled stack-up volume coil.

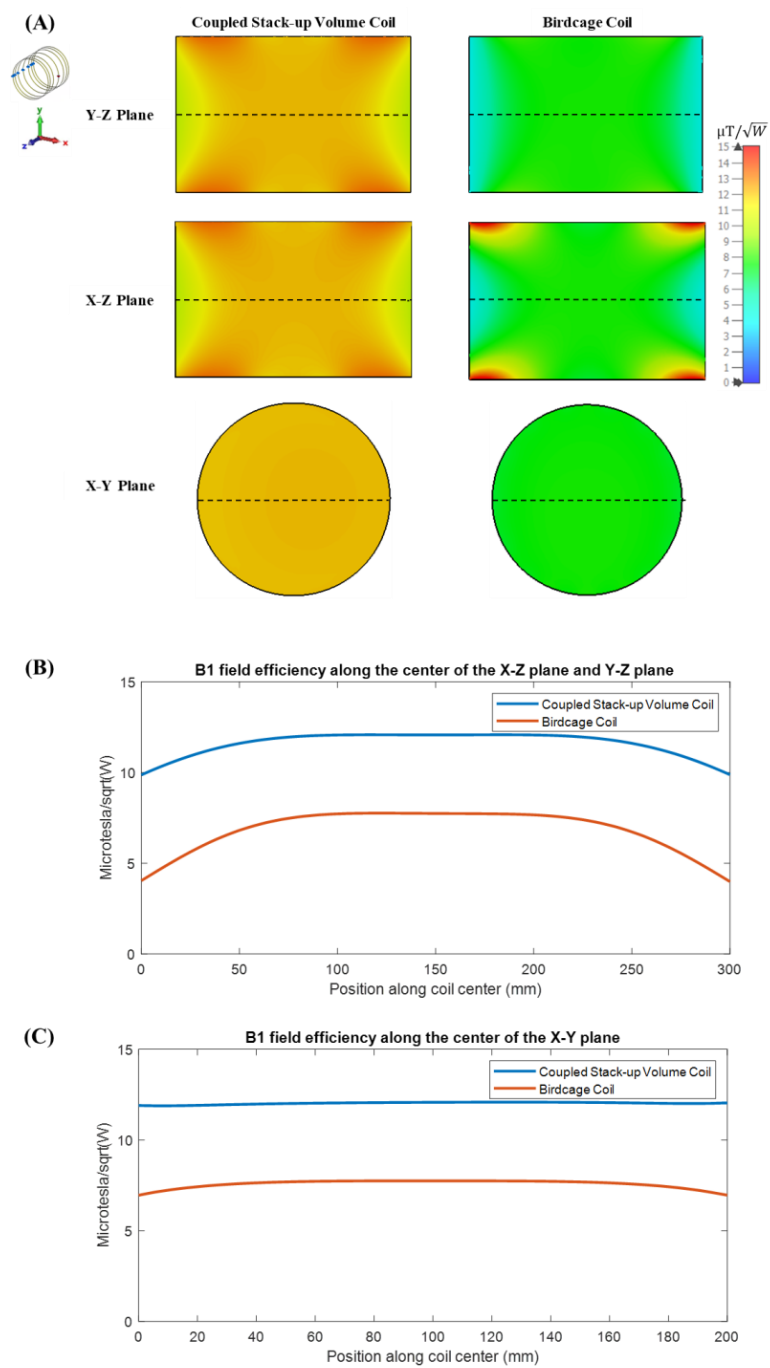

Figure 7
(A) Simulated B1 efficiency and field distribution in three orthogonal planes: Comparison between the proposed coupled stack-up volume coil and the birdcage coil loaded with a head phantom. (B)1-D profiles of the simulated B1 fields plotted along the axis of the coils, i.e. the dashed lines indicated in Y-Z plane and X-Z plane in inset (A). (C)1-D profiles of the simulated B1 fields of the coils plotted along the dashed lines shown in X-Y plane in inset (A).

|  | **Coupled Stack-up Coil** | **Birdcage Coil** |
|---|---|---|
| Average B Field Efficiency ($\mu T/\sqrt{W}$) | 9.6058 | 6.5029 |
| Standard Deviation | 0.4453 | 1.3909 |

Table 1
Simulated average B field efficiency and standard deviation inside the phantom of coupled stack-up volume coil and birdcage coil. The average B field efficiency and standard deviation are calculated by the simulation result inside the phantom. Field efficiency is collected and analyzed in a 2.5 mm step size inside the phantom.

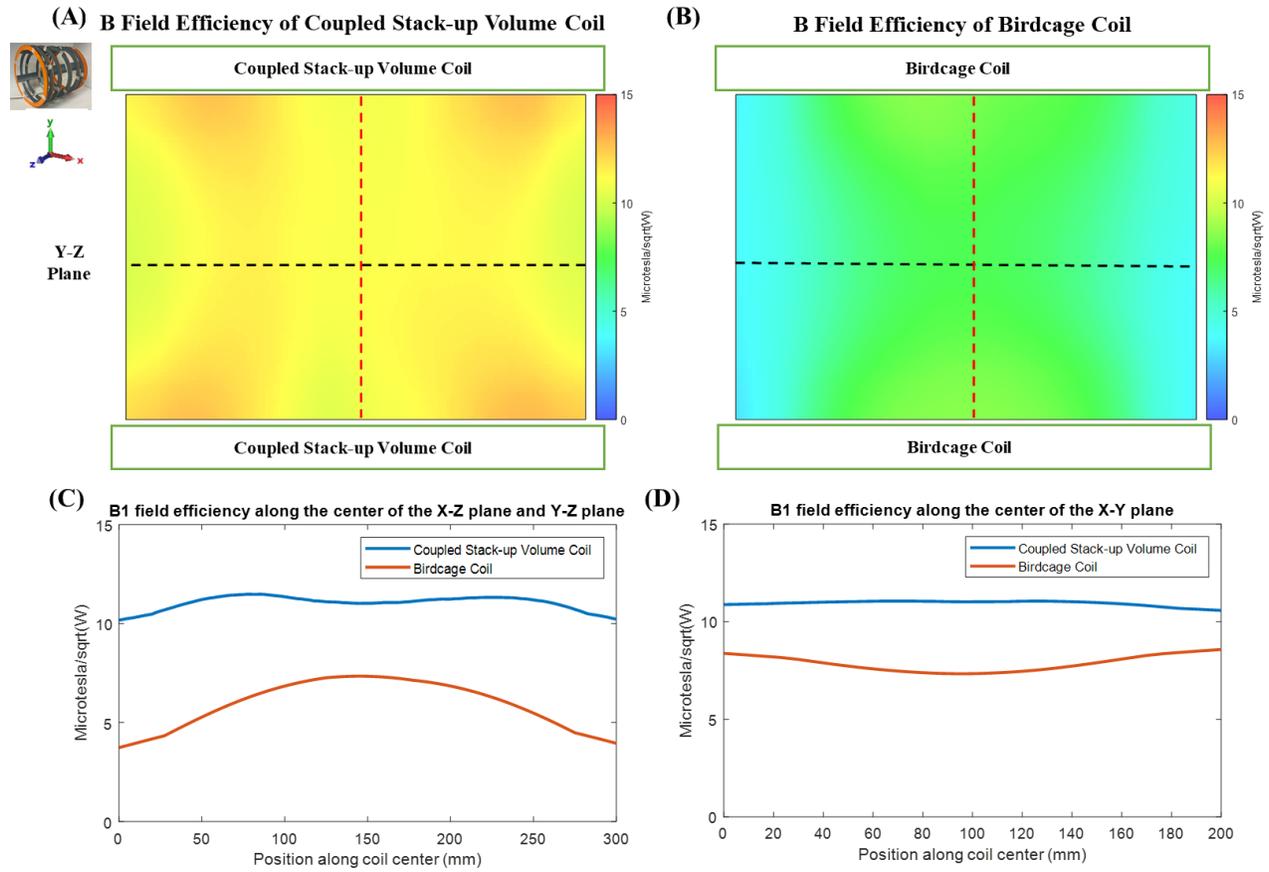

Figure 8

Measured B1 fields of the proposed coupled stack-up volume coil (A) and the same sized birdcage coil (B). 1D profiles of B1 fields of the two coils plotted along the center line of the X-Z plane and Y-Z plane (black dashed lines in (A) and (B)) are shown in (C). 1D profiles of B1 fields of the two coils plotted along the center line of the X-Y plane (red dashed lines in (A) and (B)) are shown in (D). These results demonstrate the improved B1 efficiency and homogeneity of the coupled stack-up volume coil over the birdcage coil at 0.5T.

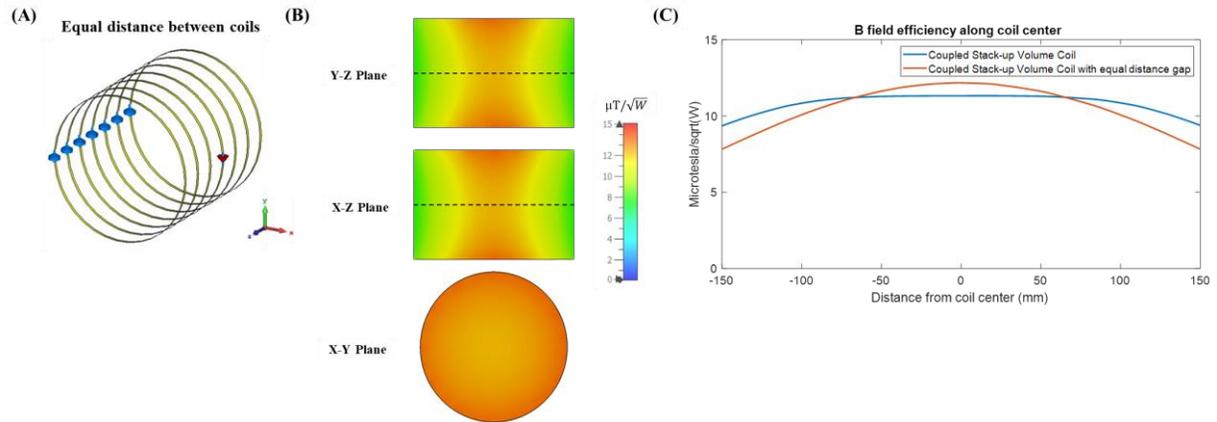

Figure 9
(A) Simulation model of equal distance coupled stack-up volume coil. (B) Simulated three planes view of the B field efficiency plot for equal distance coupled stack-up volume coil. The center of the coil has the strongest field efficiency and starts to decrease significantly when the distance to the center increases, which makes the field inhomogeneous and does not fit its function as a volume coil. (C) Simulated 1-D field efficiency comparison between coupled stack-up volume coil and equal distance gap coupled stack-up volume coil along the coil center (dash line in (B)).